# Space and social networks: A review of the literature and its implications

Steven Radil
University of Idaho
sradil@uidaho.edu

Olivier J. Walther, Ph.D.
University of Florida
owalther@ufl.edu

27 February 2019

**Abstract**

The spatial metaphor of the network along with its accompanying abstractions, such as flow, movement, and connectivity, have been central themes throughout the relational turn in human geography. However, to date networks in geography have been primarily explored either through actor-network theory or assemblage thinking, both of which embrace the network metaphor without specifically and formally interrogating networks themselves. We seek to problematize the treatment of networks in geography by exploring the largely underutilized literature on social networks as an alternative framework. Our paper discusses the conceptual connections between key concepts in geography, such as place, distance, scale, and power, and those in network theory, such as centrality, density, and homophily. Our paper opens new directions for geographers that are interested in more than the metaphor of the network.

**Keywords**

social theory, social networks, place, distance, scale, power

**Revisiting the connections between relational geography and social network theory**

  Christopher Smith introduced his 1980 review of social networks for geographers by stating that "the study of social networks is a new fad" (Smith, 1980: 500). Smith's review, one of the first about social networks specifically for an audience of geographers, argued that geography was among a wave of 'early majority' disciplinary adopters of social networks and that geographers should take them seriously as they offer a "new way of thinking about [social] problems" (Smith, 1980: 520). Nearly 40 years on, it is safe to say that social networks continue to attract interest from human geographers without becoming as widely adopted as other forms of relational thinking, such as actor-network and assemblage theories that have disseminated across the various sub-disciplines in human geography in recent years by drawing on the network metaphor.

With this in mind, our paper revisits the connections between human geographic thought and social network theory to illustrate the potential of more engagement with social networks for a now largely relational human geography. To do so, we take on three interrelated tasks. First, we discuss the relational turn in geography in order to consider social networks as part of a larger movement toward relationality in geographic thought. Second, we review a few fundamental principles of social network theory in recent geographic scholarship to showcase some of the vibrancy from this field. Lastly, we consider the conceptual connections between core geographic concepts and social networks to show that the largely underutilized literature on social networks can be as an alternative framework for relational thinking in geography. We conclude by considering the potential for further growth for social networks in geography.

**The relational turn in geography and its progeny**

Thomas Kuhn (1963) emphasized the importance of 'paradigms,' or coherent collections of claims, methodologies, and practices that serve to govern scientific inquiry. In a strict Kuhnian sense, the elevation of relational approaches in the social sciences over the last 25 years might be reasonably called a now-dominant paradigm, if not an outright paradigm shift. In geography, the so-called 'relational turn' and the importance of relations has been a core theme since the early 1990s (Müller and Schurr, 2016). A key part of this broader shift toward relations has been a reliance on the network metaphor (Marshall and Staeheli, 2015).

Networks are a form of social and spatial arrangement consisting of connected entities or elements. They have a long history in geographic thought, having been introduced well ahead of the relational turn, but without always emphasizing their social dimension. For example, in their pioneering volume Haggett and Chorley (1970) proposed different measures and models of drainage patterns, roads, subways, and pipelines while highlighting the importance of graph theory in geographical analysis. Their advocacy of graph theory, or the mathematical study of networks, also highlighted a key tension at the time: mathematical modeling proved to be fertile ground for some topics, such as transport geography, while also prompting critiques by others that such models were thoroughly divorced from any sense of social process. For example, Hadjimichalis and Hudson (2006: 859) called a graph theory approach to networks "the apotheosis of the spatial analytic approach in geography. Formal mathematical approaches to networks in geography were largely isolated to particular sub-disciplines and discounted "because of their asocial conception of social relations and spatial structure" (Hadjimichalis and Hudson, 2006: 859).

Broadly speaking, the relational turn in geography in the 1990s resuscitated the idea of the network, replacing their association with graph theoretic models with a spatial metaphor for issues of social connectivity, flows, and interactions between and within places. The relational turn refers to the increased interest across the social sciences and humanities in how social relationships and interactions constitute various outcomes and phenomena (Dépelteau, 2013) and geography's version was introduced largely through the economic geography literature (Boggs and Rantisi, 2003). A concern for social relations was manifested in key debates on structure-agency duality and the conceptualizations of scale which themselves owed much to the writings on relational space found in the influential works of Manuel Castells (1996), Gilles Deleuze and Félix Guattari (1980), Bruno Latour (1993; 2005), and others. As a consequence, the network metaphor is now routinely framed as a central concern to human geographic thought and is often now fielded in introductory texts to discuss issues of movement and/or spatial interaction (e.g., Fouberg et al., 2015).

An important element of the rebound of networks in geography is the elevation of topological understandings of space. As described by Harris (2009: 762), topology derives from "a field of mathematics studying the spatial properties of an object or network that remain true when that object is stretched." Topology has been advanced as an alternative to either absolute or relative notions of space and as a framework to better understand how human agency can at times transcend the limitations of distance, boundaries, or territory. Paasi (2011: 300-301) connected topological thinking in geography to both "the heydays of positivist thinking" associated with the quantitative revolution as well as with the theoretical efforts associated with the relational turn, such as Doreen Massey's (1993) notion of 'power geometries.' For example, John Allen's (2003) theorizations of power in geography drew on both networks and topology to suggest that space is a performed product of networked entities. Although networks and topologies are not the same, they are also often fielded together. Networks are now interpreted as social constructs that both suggest and embody the topologies that challenge more conventional notions of space.

The success of the network metaphor has also been responsible for the popularity of two relational theories in geography: actor-network theory (ANT) and assemblage theory. Both are imports that emphasize themes of connectivity, a key calling card of relational thinking, and embrace topological perspectives on space. Since Latour's (1993) influential *We Have Never Been Modern*, ANT has become a well-established approach in geography, emphasizing how "all sorts of bits and pieces (bodies, machines, and buildings, as well as documents, texts, and money) are associated together into actor-networks, configured across space and time" (Simonsen, 2004: 1334). Relations between humans and nonhumans remain the primary emphasis of ANT in geography as exploring these specific types of relations is thought to highlight the role of nonhuman entities (like the 'environment') in shaping "the conditions of the economy and the character of human culture" (Robbins, 2007: 137). Müller and Schurr (2016: 218) point out that geographers have found ANT appealing for other reasons, including understanding "the ways in which networks work on space."

Assemblage thinking emerged into geography somewhat later than ANT but has been embraced across numerous sub-disciplines. Assemblage is grounded primarily in the work of Gilles Deleuze and Félix Guattari (1980), particularly in their development of yet another metaphor, that of the rhizome. The rhizome, a botanical term for nodal parts of plant's roots system that can send out new shoots, is used to "denote a network in which, unlike in the tree-like organizations, any node can immediately connect with any other node" (Menatti, 2013: 21). This metaphor has itself been reformulated as an assemblage, a term that emphasizes "emergence, multiplicity and indeterminacy" and the "composition of diverse elements into some form of provisional socio-spatial formation" (Anderson and McFarlane, 2011: 124). It has been argued that assemblage also provides an alternative to the interrelated problems of conceptualizing agency-structure and scale in geography (De Landa, 2016).

While ANT and assemblages represent important modes of relational thinking in geography, they do not compose the entirety of possibilities. There is a vast if still largely untapped body of though on relational thought that lies adjacent to both ANT and assemblages that also relies on the network metaphor (Grabher, 2006). Here we point to the 'relational sociology' literature that has emerged out of sociology's own engagements with relational thinking (e.g., Prandini, 2015). As Emirbayer (1997: 281) put it, relational sociology is grounded in a general move away from conceiving of social reality as "consisting primarily in substances" and toward a conception of the social "in dynamic, continuous, and processual terms." Unlike relational turn in geography, this literature has not been dominated by either ANT or assemblage but has emphasized the idea of the 'social network' or the "particular figurations of social ties" (Emirbayer, 1997: 298).

**Social networks**

Research on social networks emerged in the 1930s on theoretical foundations inspired by German sociologist Georg Simmel and British anthropologist Radcliffe-Brown. Until the 1970s, however, none of the schools of thought that developed in sociology, anthropology and psychology, both in the United States and Great Britain, succeeded in building a universally accepted paradigm within the social sciences. The work of Harrison White and his students (e.g., White et al., 1976) contributed to make network analysis a field of research in its own right. Research on networks experienced a sudden boom at the end of the 1990s, due to the greater availability of relational data and by advances in computational science.

Today, research on social networks remains largely composed of two communities: that of the social sciences, whose primary interest is to use network theory to better understand social structures, and that of the physical sciences, for which the primary interest is to develop ever more rapid algorithms that can deal with networks composed of millions of actors. We note with some irony that this enduring division within network research is no small paradox for scholars supposedly working on the relations between communities. In any event, it is naturally with the first perspective – that of *social* networks – that synergies with geography are most numerous.

Within the social science tradition, social network research draws on a set of theories designed to illustrate the processes at work within networks, such as the implications of being the most prominent actor in a network, and on sets of measures to represent, analyze and model these ideas, such as centrality. Accordingly, social network theory seeks to clarify the "processes that interact with network structures to yield certain outcomes for individuals and groups" (Borgatti and Halgin, 2011: 1168). This typically involves identifying the importance of the 'global' structure of the network for the exercise of agency by individual actors. This approach can be thought of as a 'network level' of analysis which seeks to characterize the form of the networks and its implications. For instance, centralized networks, such as the so-called star network where every actor is connected to a central actor but otherwise unconnected to each other, are theoretically more effective in coordinating small group activities than are decentralized networks in which the flow of information and resources is more uncertain.

Social network research is also concerned with individuals within networks. This approach involves understanding how an actor is linked to the rest of the network and what strategies they develop to exercise their agency (Brass and Krackhardt, 2012). Being 'central' in a set of relations is an important concern here. Some actors are central because they are 'embedded' within a tightly knit community of friends, kin or allies or because they are connected to many well-connected actors. Others actors, called brokers, are central because they bridge communities that would otherwise be disconnected (Burt, 2005). Research in a variety of disciplines and geographical settings, from patent co-authorship networks to organizational networks and terrorist networks, has shown that social capital results from a combination of embeddedness and brokerage (Fleming et al., 2007: Everton, 2012). Successful, innovative or resilient actors are ones that are simultaneously well integrated into a dense group of close relationships and able to create contacts beyond their own community.

Social network theory considers social ties as possible conduits for flows of resources or as bonds that contribute to social homogeneity (Borgatti and Halgin, 2011). On the one hand, the "flow tradition" investigates why some network actors seem to be better off, get promoted earlier, are more innovative, or make more profit than others. In so doing, these social capital theories describe how networks allow some well-placed individuals to access new resources and ideas or to better coordinate collective activities. Granovetter's (1983) theory that states that people find jobs through "weak ties" that provide non-redundant information or Burt's (2005) theory of structural holes that argues that brokers do better because they have a competitive advantage in seeing and developing good ideas are two classical examples of such theories. On the other hand, the "social homogeneity" tradition also addresses the question of why certain social actors tend to resemble one another and form homogeneous communities. Barry Wellman's (1979) study of how residents of East York, Canada, were developing ties that were no longer confined to their neighborhood is representative of such approach. In recent years, social network theory has also been widely applied to explain the process of contagion that leads certain individuals, but not all, to start smoking, using drugs, or sharing political ideas (Eveland et al., 2018; Mason et al., 2016).

**Synergies between social networks and geography**

In the last two decades of the relational turn, the idea of combining geographical and network theory has gained significant attention. Given this resurgent interest in social networks in human geography, it is clear that social networks can be fielded to better understand some of the most fundamental concepts in geography. To illustrate that potential, we have selected four such concepts, place, distance, scale, and power, to discuss how they align with social network theory. Along the way, we also point to important works by geographers that have drawn on such efforts at synthesis. The selected concepts have been chosen to highlight some of the key intersections between space and social networks in a manner which may be instructive to both geographers and network scientists.

Figure 1 below serves as an overview of the basic connections between networks and these geographic concepts. For instance, when a social network is projected on geographical space or across a study region, the most prominent information is the location of the various actors that compose the network. Each actor of a social network, represented as nodes in the figure, develops and maintains affective relationships, identities, and attachments to the places in which they are primarily located, indicated with vertical dashed lines. In typical geographical fashion, actor location can always be measured in absolute terms through geographical coordinate pairs and characterized or contextualized by the attributes of the places with which they are associated.

Assuming the location of any two actors is not identical, several forms of distance can also be measured between nodes and their associated places to reflect a variety of conceptualizations of proximity in Euclidian or other types of space, including social, institutional, cognitive or organizational. Such distance measures, shown by the dotted line between places in Figure 1, would always be measured between actor locations (places) but not necessarily follow the form of an existing network. This information can be used to assess or condition the possible formation of new interactions between actors over time. In a general sense, actors that are close in geographic space may be more likely to form relations but those that are close on multiple social dimensions may also heighten that potential for interaction.

Geographical scale is a contested concept to be sure but most conventionally refers to the geographic scope or reach of social processes, authority, or jurisdictions. These processes may only involve small distances (generally referred to as the local), or vast distances (the global), distances in between, or often at multiple scales simultaneously. Scales and the processes behind them also often result in territorially-defined regions of different areal extents, such as states, provinces, counties, municipalities, or neighborhoods which networks may (or may not) span. Connected to this interpretation is the notion of boundaries and borders that are associated with such delineated spaces and the potential for such features to impact interaction. Networks may also be implicated in the construction of such scalar and territorial divisions, shaping the social processes that produce such geographies.

Finally, power is both an input to and a consequence of the unequal distribution of relations between social actors. Places, distance and scales all affect the ability of actors to access resources and develop strategies to achieve their goals. Similarly, networks can recursively facilitate how power works geographically while also shaping how networks develop within places and across distance and scales.

Figure 1. Social networks and space

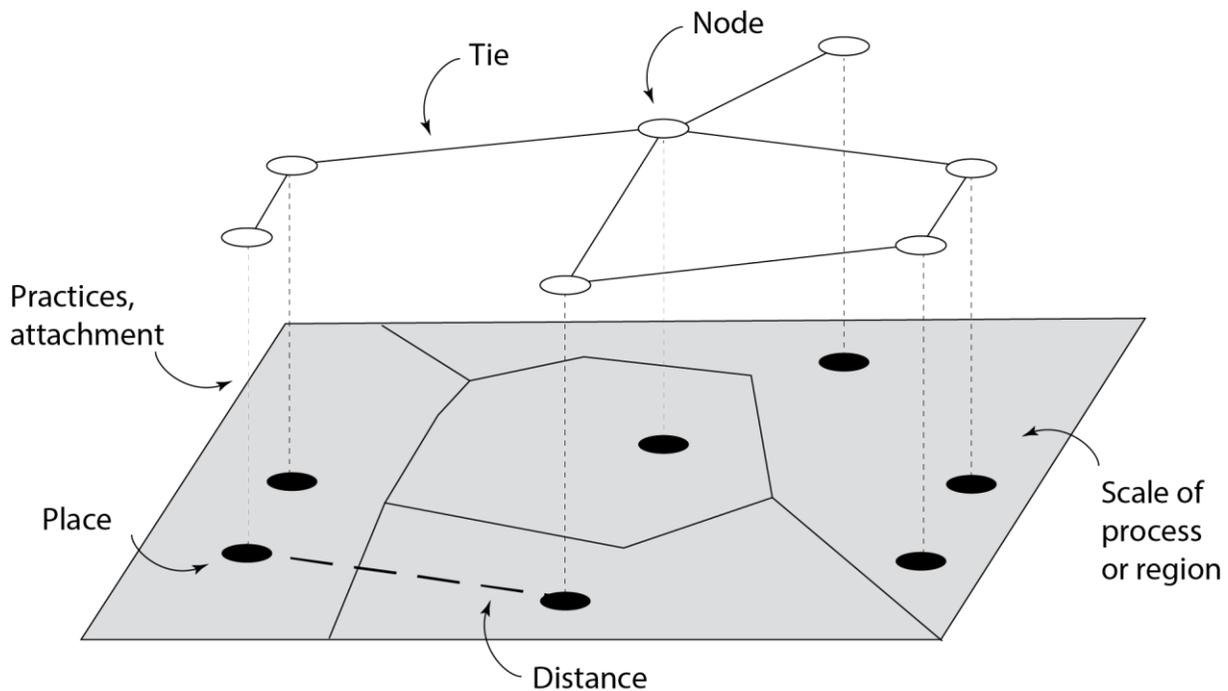

Source: Adapted with permission from Sohn et al. (2010).

Table 1 below provides an overview of the more detailed discussions about these concepts that follow. While places are conceived as a series of unique localized settings connected to others or made through people's actions in the geographic tradition, social network theory looks at places through the concepts of centrality and community. Whereas geographers have developed several measures and forms of proximity, network scientists rely on the concept of homophily to study how geographical distance affects the probability of sharing common attributes, or homophily. Similarities between the two disciplines can also be observed when scales are concerned: while geography has debated conceptualizations of vertical and horizontal scales, network theory speaks of embeddedness and clustering. Finally, geographers and network scientists have helpfully developed a common understanding of power, as geography and networks are understood as both the results of and mediators for the exercise of power.

The discussions that follow are not meant to be definitive or to represent mutually exclusive categorizations but rather to suggest possibilities for generating more geographical engagement with social networks.

**Table 1. Correspondence between geographical and network concepts**

| **Geographic concept** | **Geographic traditions** | **Social network theory** |
|---|---|---|
| Place | A) Sets of places: a series of unique localized settings connected to others<br>B) Place-making: a unique localized setting made through people's actions | A) Centrality: assessing how important a node is in a network<br><br>B) Community: identifying groups of densely connected nodes in a network |
| Distance | Proximity: nearness in physical, functional, cognitive and institutional space | Homophily: the tendency to bond with similar others |
| Scale | A) 'Vertical' scale: hierarchical 'levels' of territorially defined spaces<br>B) 'Horizontal' scale: size or scope of geographic process or outcome | A) Embeddedness: overall position of actor in a set of relations<br><br>B) Cluster: a community bounded by social, economic, religious or political boundaries |
| Power | Relational power: the ability to reach people and places across scales and distance | Relational power: the ability to access and control resources through network flows |

Source: authors

*Networks and place*

Place is one of the most important and complex concepts in human geography. While the literature on place is too voluminous to recount here, we rely on Staeheli's (2003: 159) distinction between place as an outcome and place as a process to consider how networks are implicated in place-based research. In the first case, places are understood as the unique localized setting through which human agency unfolds, with an emphasis on how a place may impact people and their choices. The second case considers how places come to be the way that they are, with an emphasis on the processes of place-making or how people impact a place. We call the first tradition 'places in networks' to stress how places are connected to one another and the implications of these arrangements, and the second 'networks in place', which focuses on how social relations at various scales help to constitute a place. Social networks intersect with both place traditions.

Research around 'places in networks' tends to draw on a conceptualization of place that has been critiqued at times but remains widely employed: place as small-scale regional space that is at least partly defined through absolute location. This tradition employs a largely descriptive approach to place or what Cresswell (2004: 51) called "the common sense idea of the world being a set of places each of which can be studied as a unique and particular entity." Places are formulated as pre-existing discrete entities that are differently located in geographic space and possess a unique mix of both social and physical characteristics. Networks are often fielded in this tradition to examine how some set of places are materially connected to one another to partly explain the characteristics of the places or to characterize the broader impacts of these connections. This largely happens by imagining places as nodes that are connected by various means to form a network that spans the space in between.

The clearest example of 'places in networks' is found in Peter Taylor's "Globalization and World Cities" project, which emphasizes the need to consider the economic relations of cities to one another, which he calls 'inter-city relations'. Taylor (2004: 93) explicitly used network concepts to identify certain cities as "important strategic nodes" that hold more prominent positions in the overall 'world city' network. This highlights the network concept of centrality, or the idea that some actors in a network are connected in such a way to yield benefits to themselves or to otherwise increase their status, influence, or power. Although there are many different interpretations of and ways to measure centrality, Taylor and colleagues (2004) relied on two in particular: the idea that the most central actor has the most ties to other actors (degree centrality) and the idea that the most central actor lies in between otherwise disconnected parts of the network (betweenness centrality). That much of the subsequent scholarship on global city networks has emphasized alternate measures of interaction between places underscores that having a central position in a network has implications for a place remains relatively accepted.

Research involving 'networks in place' emphasizes the ways places are always being remade over time through people's actions, and that places are always situated within multi-scalar processes that are also dynamic (e.g., Massey, 1994). In contrast with the previous discussion, the way in which networks are implicated in place-making is the point of emphasis in this tradition. Rather than view a place as a pre-given node in a larger network of relations that connects it to other places, the dynamics of a place is interrogated by considering the networks of social relations that help to make it the way it is. Although the past always matters for the realization of a place at any given moment in time, by focusing on the 'networks in place,' a place is never considered a pre-given entity in this approach.

Because placemaking occupies such an important role in geographic thought, there are many examples of 'networks in place' across multiple subdisciplines in human geography. Far from being limited to the study of economic spaces, this approach has been particularly fruitful to explain governance and social issues, including the way in which social movements develop, change, and spread (e.g., Nicholls and Uitermark, 2017). For instance, Miller's (2000: 101) work on anti-nuclear networks, shows that each activist group grew its membership by connections through "personal friendship and acquaintance networks" and other local civic organizations. This highlights the importance of network clusters, or collections of actors with relatively dense relations internally and sparse relations externally, in understanding how the activist networks grew and mobilized within places. Miller (2000: 113) also notes that such social clustering is essential for identity formation within places as the dense relations yielded a sense of community among activists that was highly localized. Lastly, Miller identified a connection between place attachment and network clustering as groups with a lower sense of place attachment were those that perceived a large gulf between their efforts and their satisfaction with the political situation in their city. In other words, the efficacy of placemaking, even among densely connected clusters of actors, is geographically uneven. Networks matter for places and places matter for networks.

*Networks and distance*

Spatial proximity is well known for increasing the likelihood of developing and maintaining social ties. Spatially close individuals have a higher probability to have contact with each other than distant ones, which, in turn, enhances individual integration, cohesion, shared values and innovation (Hipp et al., 2012). Nowhere in our discipline has the link between distance and social networks been more thoroughly investigated than in economic geography, where major conceptual and methodological advances have been made since the mid-2000s (Grabher, 2006). The introduction and development of network analysis within economic geography has led to the publication of books (Bathelt and Glückler, 2011; Glückler et al., 2017) and special issues (Chilvers and Evans, 2009; Adams et al., 2011; Glückler and Doreian, 2016), often with network scientists.

In addition to a certain affinity for mathematical models, the attention paid to the economic dimension of networks owes much to the evolution of the main objects of analysis of economic geographers – the entrepreneur, the firm, the industrial cluster and the city. Network analysis has now been widely applied to key topics related to the geography of innovation and knowledge, such as industrial clusters, inter-firm, intra-organizational and community-based relations, global value chains, global production networks, world city network and economic migration.

While Western economies are increasingly turning to knowledge-intensive services and technologies, the use and value of social networks within and between businesses is growing. In this knowledge economy, Euclidian distance is not a sufficient condition to explain the formation of networks and for localized knowledge spillovers to develop and must be supplemented with other forms of proximity (Boschma, 2005). In addition to being close to each other, entrepreneurs and firms also thrive because they share the same knowledge base and expertise, have established relationships of trust, and belong to similar institutional settings.

The benefits of physical, cognitive, organizational, social and institutional proximity are particularly valuable for the exchange of non-codified information that is deeply rooted into social or professional contexts and can only travel with great difficulty (Grosser et al., 2010). Because non-codified information is widely regarded as one of the main sources of long-term innovation (Howells, 2012), geographers have paid great attention to the value of face-to-face communication in business communities. Geographic proximity encourages social actors to meet and socialize, notably through chance encounter in large and heterogeneous urban cores. Proximity networks explain the revival, or the long-lasting attraction of many city centers who have succeeded in attracting an increasing share of knowledge-intensive activities, such as the "City" in financial centers (Neal, 2012). This trend has taken place despite the decrease of transport cost and the improvement of communication technologies which essentially favor the transmission of codified information.

This renewed interest for various forms of proximity has led economic geographers to innovate both geographically, conceptually and methodologically. Many economic geographers increasingly turn to the question of how the structure of an entire network influences collective outcomes (see Doreian and Conti, 2012; Broekel, 2015; Glückler and Panitz, 2016) rather than on the how the position of an actor can influence individual behavior through dyadic (Broekel and Boschma, 2012) or triadic formation (Ter Wal, 2014). By moving away from the characteristics of individual actors and increasingly considering the general properties of whole networks (Glückler and Doreian, 2016), geographers contribute to bridge the two fundamental dimensions of network science.

Conceptually, economic geography has also moved away from a static conception of network to a more dynamic understanding and modeling of social structures (Glückler, 2007). Encouraged by the evolutionary economic geography approach, recent studies have focused on the role of space in the long-term technological evolution of industrial clusters and its effects on regional growth and innovation (Balland et al., 2015; Boschma et al., 2014). The factors that drive the formation, persistence and dissolution of social ties are subject to particular attention, as they may explain the trajectories of industrial clusters and regions (Juhász and Lengyel, 2018). Some of the conclusions of these studies mirror those of network science. For example, the formation of informal knowledge networks is now widely seen by geographers as a complex outcome of embeddedness and brokerage (Balland et al., 2016; Breschi and Lenzi, 2016; Broekel and Muller, 2018). While embeddedness in a dense network of peers provides trust and reputation at the local level, brokerage ties to more distant partners provide non-redundant information that enhance innovation.

The study of increasingly global and dynamic social networks has forced economic geographers to develop new methodological tools that can take into account the spatial and temporal evolution of networks (Broekel et al., 2014). Among these tools borrowed from other disciplines is blockmodeling, a projection of a larger network that can delineate underlying structures (Glückler and Panitz, 2016). Blockmodeling reorganizes network data in such a way that the structural position and roles of the people involved in networks and the overall division of labor between them is highlighted. Another tool is multi-level analysis, an approach that allows to study the contribution of several types of actors, such as individuals or organizations to networks. Multi-level analysis is particularly adapted to capture the role of temporary proximity for business relations during trade fairs (Bathelt et al., 2014; Maskell, 2014).

In addition to being a key determinant of economic activities, spatial proximity also influences social behaviors. This is particularly evident for crime and substance abuse, where geographical approaches have generally considered both network and spatial issues simultaneously. For instance, while Papachristos (2009) considered the overall network position of individual gang members that carried out violent acts, Radil et al. (2010) examined the twin impacts of networks of gang rivalries and territorial competition on the production of spatial patterns of urban violence. Violence, the authors show, does not take place just between geographically proximate gangs but also between gangs who are topologically close, i.e. who share a similar structural relation within the overall network of urban violence. Similarly, Schaefer (2012) found that both spatial and social network effects helped explain groups of young people that committed crimes together. Political geographers have also utilized network approaches. For instance, Flint et al. (2009) used networks to model the diffusion of World War I and Radil and Flint (2013) examined the creation of regional interstate security networks in Africa as a consequence of ending the civil war within the Democratic Republic of the Congo.

A similar interest for the study of distance can be found in the network community, where geographical proximity is often related to the probability of sharing common social attributes, or homophily in network jargon (McPherson et al., 2001). While most network scientists agree that physical distance affects the probability of social ties, it is nevertheless less clear how such a relation can be formalized, probably because different kinds of ties are affected in various ways by distance. Preciado et al. (2012: 13) argue that the existence and creation of friendships can be approximated by a linear function of log-distance. Studying face-to-face and telephone interactions before the Internet, Mok et al. (2007) find that influence does not decrease regularly with distance but follows different thresholds. Working on U.S. and international citations of patents held by biotechnology firms, Tallman and Phene (2007) find that knowledge flows are not inversely related to distance of separation but follow a non-linear curve. Studying mobile phone networks in Europe, Onnela et al. (2011) find that network ties decay with distance but tie strength is nearly flat with distance, suggesting that strong ties between people are relatively indifferent to separation.

Recent research in both geography and network science shows that social networks do not only decline rapidly with distance but are also very sensitive to cultural, linguistic, political and physical boundaries (Skillicorn et al., 2018). Boundaries introduce various distortions that are similar to adding an extra distance between social actors. Comparing co-inventors networks in Europe and the United States, Cerina et al. (2014) show for example that connectivity decays exponentially in Europe due to the existence of national communities of inventors, and declines as a power of distance in the borderless market of the US. Boundaries limit international trade flows and subsist even after regional agreements to facilitate regional integration are put in place. They also constrain the mobility of labor between countries, particularly when workers must cross a linguistic border (Walther and Reitel, 2013). Invisible boundaries between co-workers also shape the formation and structure of intra-organizational networks (Sailer and McCulloh, 2012).

The effect of borders on social networks seems to persist regardless of the technologies used to communicate. Thus, the increasing use of email and mobile phones do not seem to have profoundly altered the effect that distance and borders exert on communications (Takhteyev et al., 2012). In addition to hindering the mobility of goods and people, boundaries also constrain political violence, which often occurs at the neighborhood or sub-national level. The turf of the gangs studied by Radil et al. (2010) in the Hollenbeck Policing Area, for example, is strongly constrained by the Pasadena Freeway and the Los Angeles River in the west, two elements of the urban landscape that tend to limit interactions with other gangs.

*Networks and scale*

In its most basic form, scale typically refers to the geographic extent or size of some phenomena or process. And yet, scale has been the subject of a great deal of theoretical debate as to its precise meaning and usage in human geography. The traditional interpretation of scale is one of a nested hierarchy of differently sized spaces (or 'levels') that are often territorially defined and range from the local, the national, the regional, and so on through the global. These scales are frequently treated as observational and analytical frameworks defined by the relations between people and the various political and economic institutions that shape daily life (Dahlman, 2009). This 'vertical' conceptualization of scale has been the focus of a great deal of scholarship in geography. Key themes involve interrogating which scales are more important for certain issues, introducing new scales (such as the body or household), considering how activity at one scale impacts others, and exploring how actors actively navigate or even create scales (see Jones et al., 2011 for a thorough review).

A vertical sense of scale has also been roundly challenged, with some arguing in favor of a 'horizontal' or 'topological' sense of scale where attention is paid to how social processes manifest in different spatial configurations which may or may not match neatly with other pre-existing configurations of economic activity or governance. These approaches, driven largely by investigations of globalization, emphasize the need to work through the processes that span territorial boundaries in order to "link places together in the world" (Aoyama et al., 2011: 127). It is through the horizontal or topological sense of scale that networks have become a central metaphor. Interestingly, the move to a horizontal sense of scale was part of the larger impact of actor-network theory in geography. For instance, Latour (1996: 370) called for a networked view of scale as "fibrous [and] thread-like," a notion that quickly informed others in geography. Leitner (2004: 237) advocated for scale as "networks that span space rather than covering it, transgressing the boundaries that separate and define these [territorial] political entities".

Scale is rarely directly addressed directly in the social network literature, particularly in the vertical sense. Of course, territorially bounded spaces are often used uncritically in many social sciences as presumed 'containers' for a process under investigation and studies that use such scales to delimit or bound social networks are common. For example, Hess (2004) coined the term 'overterritorialized' to warn against overly privileging certain scales when adapting social networks for economic geography. He advocated for the idea that actors are embedded in multiple types of networks, some constrained spatially at certain scales, others stretching beyond. Embeddedness, itself a key social network concept, refers to how one's overall position or standing in a set of social relations shapes behavior (Granovetter ,1985). Following Hess then, a concern for vertical scale would then also necessarily involve assessing the embeddedness of actors both within and beyond their immediate settings. Another key concern of vertical scale is how the actions of people and groups cross, jump, or link scales. Social networks can provide insight into this process as well. Radil and Flint (2013) observed how peacemaking efforts at the nation-state level led to new regional networks of interstate cooperation on security issues.

      Horizontal scale is also an infrequent concern as the spatial size or scope of a network is often not pre-given, particularly when concerning the behaviors of individuals in a social network. The social networks of individuals are typically sampled or observed by tracing relations outward from a single focal individual (an egocentric network or egonet). While there are many specific questions of interest regarding egocentric networks, the larger concerns have to do with the idea of community and how this varies relationally among members. As described by Chua and Wellman (2011: 236), "rather than treating community as spatially bounded units... egocentric networks treat community as networks of social relations emanating from a focal individual." Such an approach has the advantage of allowing the horizontal scale of a network to emerge out of the relations in question offering a key alignment with the scale literature in geography. Instead of dividing space and societies according to predetermined categories such as regions, race, or class and studying the characteristics of each, social network scientists consider all stakeholders involved in a particular event or domain, map the links between them and, only then, determine how the network is split internally according to spatial or social lines.

      Community then, can be seen as the clustering of dense relations within sub-areas of a network even when individual relations may stretch over significant geographic distances. Comparing the relative density of relations between egonets has been a primary approach in the social network literature (e.g., Wellman, 1979). However, geographers have begun to add their own interpretation to issues of clustering. For instance, Frei et al. (2009) consider the relationships between the spatial extent of egonets with patterns of mobility finding connections between dispersed community relations and social standing within places. Although a relatively new direction of inquiry, this holds promise as it aligns well with the notion of horizontal scales emerging from the actions of people.

*Networks and power*

The relational turn in geography has profoundly reconfigured debates on power. Few geographers today would claim that power only resides in the possession of certain mineral, demographic or military resources or in the behavioral tactics adopted by people or countries to gain influence. Geography has developed an understanding of power that is progressively liberated both from the idea of territorial containment and spatial contiguity (Allen 2003; 2010). Accordingly, power is not a thing that can be possessed. Rather, it only exists as an emergent effect *between* actors and as long as the actors are involved in a collaborative or conflictual relationship. This has been an important conceptual shift in geography as individuals, social movements, corporations, cities, and states can all be seen as powerful despite imbalances in their relative material means.

For relational geographers, the skepticism for methodological individualism is grounded in the fact that social actors are embedded in social and cultural norms. Individuals, communities, firms, regions and countries are dependent on one another and as such should be seen as nodes within a network that is greater than the sum of its parts. New properties, such as power, emerge not necessarily because of the people or the spatial units themselves, but because of the ways in which they are connected.

The relational approach developed in geography to understand power relies heavily on the concept of network. As Allen (2010: 2898) argues, power "is not something that circulates or flows in networks, it is an effect of the social interactions that hold the networks together". Nowhere is this more evident than in metropolitan regions, where representatives from local municipalities, regional bodies and state agencies build policy networks across institutional levels and administrative boundaries (Walther and Reitel, 2013). Their power comes from their ability to reach other actors, build consensus and form coalitions that coordinate activities around a particular policy event.

The relational conception of power shared by geographers is remarkably similar to that developed in network science. In contrast to individualistic or atomistic approaches that focuses exclusively on the attributes of the actors, one of the fundamental assumptions of social network theory is that social life is created primarily by relations and the patterns formed by these relations. Therefore, the *relations* defining a position in any social structure - not the *attributes* of the actors – are the direct cause of the observed outcomes (Brass and Krackhardt, 2012). This has major consequences for the study of power, which is seen in network theory as composed relationally through the interactions of the different actors involved. Actors can't be powerful per se. They are in a powerful structural position that allows them to *access* resources that would otherwise be out of reach and to *control* resources that could be exploited by other actors (Smith et al., 2014).

The distinction between power as access and power as control is critical when it comes to measure power relationally. When power is seen as a mean to gain *access* to resources, degree, closeness or eigenvector centrality can assess to what extent one particular actor is densely connected, close to the most central actors or connected to structurally important actors. When power is defined as a mean to gain *control*, betweenness centrality can tell which actors play the role of brokers. These centrality measures work well when a network only contains positive ties, as in a network of friends, but reach their limits in a politically charged network in which actors dislike, hate or fight each other. Being connected to many other actors suddenly becomes a liability.

Recent developments in network science allow to take into account alliances and adversarial ties simultaneously. The Political Independence Index developed by Smith et al. (2014), for example, measures to what extent actors are structurally independent from other actors for resources and support. Applied to the field of international relations, these principles have contributed to reassess the importance of alliances between countries. If, generally speaking, states have interest in having many allies and few enemies, the choice of their allies and enemies matters greatly from a structural perspective. Being connected to an ally who is himself free of threat reduces autonomy, while being connected to an ally who is tied to many enemies offers greater autonomy.

So far, few geographers have ventured to spatialize power using the concepts and mathematical developments introduced by network scientists. The application of several measures of power that can capture positive and negative ties within an entire network seems however very promising in geography if one wants to go beyond classical approaches, such as autocorrelation, that look at spatial contiguity between actors. In addition to the field of international relations, recent advances in network science can be usefully applied to the study of modern conflicts, one of the particularities of which is to bring a large number of state and non-state actors together and to cross borders, mixing the social and the spatial in new configurations.

**Conclusions**

The ascendancy of relational theory in human geography has created new opportunities for us to reconsider social networks. For example, in trying to introduce social networks to geographers in 1980, Smith listed several network concepts important at the time (directionality, homogeneity, connectivity to name a few) and strove to clarify why geographers might find them useful. In this article, we have tried to reverse this approach by instead discussing core geographic concepts to clarify why social networks help us better understand them.

This reversal is largely only possible because of the dominance of the network metaphor in the most recent wave of geographic theory which has led to a relational rethinking of space, place, scale, and power. Similarities between relational geographic thought and social network theory has been built on a common interest for the study of places, seen both as a node in a network and as a localized network itself; on shared concerns in different types of distance and barriers to distance; on recent conceptual developments that test the validity of preconceived scales; and on a definition of power that relies on relations rather than attributes.

If relational thinking has reached a saturation point in human geography, then what more is needed for social networks to be fully integrated into geography? Smith (1980: 500) himself listed some of the steps by which topics, issues, or concepts can become 'institutionalized' within a discipline's core; these include timely review articles, special issues of important journals, readers and textbooks, special sessions at professional meetings, and the creation of a formal organizational subdivision. It is helpful to revisit this list as social networks in geography currently meet relatively few of these criteria. For example, there are no textbooks or readers on social networks in geography written either by or for geographers. Nor has there been a clearly established research community around social networks in geography. The only one of Smith's criteria that seems to be currently met is the presence of several social network-related sessions at the annual American Association of Geographers conference.

Clearly more is needed for the status quo around social networks in geography to change. With this in mind, we have three suggestions to better bridge the gap between our paper and the rest of Smith's criteria. First, we agree that more collections of papers that emphasize social networks would be welcome but stress that these cannot simply focus on implementing some version of social network analysis per se. This would only serve to relegate interest in social networks to issues of methodology and run the risk of marginalizing social networks as merely a set of quantitative tools. This goes against the general contemporary trend in the social network literature, which is to integrate both qualitative and quantitative data to understand the temporal and geographical evolution of networks, the formation of ties, and their intensity. Instead, papers that are truly interested in social networks should also strive to demonstrate linkages between geographic theory and social network concepts. In other words, we need more explicitly geographic theorizations of social networks.

Second, textbooks and readers that illustrate and demonstrate issues of both social network theory and methods for human geographers are clearly needed. Related to this are a need for the development and distribution of teaching and curricular materials. Like for so many issues in geography, interest in social networks can be first realized by encountering ideas in the classroom. However, the extent to which social networks are actually taught by geographers and in geography programs is unclear. While we are aware of three current examples in US-based geography departments, each of these introduce social networks as subpart of a Geographical Information System (GIS) course. This tends to reinforce a presentation of social networks as tools rather than a set of theories and concepts that directly align with some of geography's most fundamentals assumptions about place, distance, scale, and power.

This concern is connected to our third, and final recommendation, which is to call for the development of a formal sub-community of scholars in geography interested in social networks. It is with irony that we note that those interested in themes of relations and connectivity remain somewhat disconnected from each other. This is likely partly a function of pre-existing sub-disciplinary categories in human geography and their associated publication outlets serving to limit the visibility of work done using social networks elsewhere. The time may be now right to look for alternate organizational outcomes (and perhaps for some selectively positioned brokers) to better integrate such subcommunities.

**Bibliography**

Adams J, Faust K and Lovasi GS (2012) Capturing context: Integrating spatial and social network analyses. *Social Networks* 34(1): 1-5.

Allen J (2003) *Lost Geographies of Power*. Oxford: Blackwell.

Allen J (2010) Powerful city networks: more than connections, less than domination and control. *Urban Studies* 47(13): 2895–2911.

Anderson B and McFarlane C (2011) Assemblage and geography. *Area* 43(2): 124–127.

Aoyama Y, Murphy J and Hanson S (2011) *Key Concepts in Economic Geography*. London: Sage.

Balland PA, Boschma R and Frenken K (2015) Proximity and innovation: From statics to dynamics. *Regional Studies* 49(6): 907-920.

Balland PA, Belso-Martínez JA and Morrison A (2016) The dynamics of technical and business knowledge networks in industrial clusters: Embeddedness, status, or proximity? *Economic Geography* 92(1): 35-60.

Bathelt H and Glückler J (2011) *The Relational Economy: Geographies of Knowing and Learning*. Oxford: Oxford University Press.

Bathelt H, Golfetto F and Rinallo D (2014) *Trade shows in the globalizing knowledge economy*. Oxford, Oxford University Press.

Boggs JS and Rantisi NM (2003) The 'relational turn' in economic geography. *Journal of Economic Geography* 3(2): 109-116.

Borgatti SP and Halgin DS (2011) On network theory. *Organization Science* 22(5): 1168-1181.

Boschma R (2005) Proximity and innovation: a critical assessment. *Regional Studies* 39(1): 61–74.

Boschma R, Balland PA and de Vaan M (2014) The formation of economic networks: A proximity approach. In: Torre A and Wallet F (eds) *Regional Development and Proximity Relations*. Cheltenham, UK: Edward Elgar, pp. 243-266.

Brass DJ and Krackhardt DM (2012) Power, politics, and social networks in organizations. In: Ferris GR and Treadway DC (eds) *Politics in Organizations: Theory and Research Consideration*. New York: Routledge, pp. 355–375.

Breschi S and Lenzi C (2016) Co-invention networks and inventive productivity in US cities. *Journal of Urban Economics* 92: 66-75.

Broekel T (2015) The co-evolution of proximities–a network level study. *Regional Studies* 49(6): 921-935.

Broekel T, Balland PA, Burger M and van Oort F (2014). Modeling knowledge networks in economic geography: a discussion of four methods. *The Annals of Regional Science* 53(2): 423-452.

Broekel T and Boschma R (2011) Knowledge networks in the Dutch aviation industry: the proximity paradox. *Journal of Economic Geography* 12(2): 409-433.

Broekel T and Mueller W (2018) Critical links in knowledge networks–What about proximities and gatekeeper organisations? *Industry and Innovation* 25(10): 919-939.

Burt R (2005) *Brokerage and Closure*. Oxford: Oxford University Press.

Castells M (1996) *The Rise of the Network Society*. Oxford: Blackwell.

Cerina F, Chessa A, Pammolli F and Riccaboni M (2014) Network communities within and across borders. *Nature* 4(4546): 1–6.

Chilvers J and Evans J (2009) Understanding networks at the science-policy interface. *Geoforum* 40(3): 355-362.

Chua V and Wellman B (2011). Egocentric networks. In: Barnett G (ed) *Encyclopedia of Social Networks*. Los Angeles: Sage, pp. 237-238.

Cresswell T (2004) *Place: An Introduction*. Malden, MA: Blackwell.

Dahlman CT (2009). Scale. In: Gallaher C, Dahlman CT, Gilmartin M, Mountz A and Shirlow P. *Key Concepts in Political Geography*. London: Sage, pp. 189-197

De Landa M (2016) *Assemblage Theory*. Edinburgh: Edinburgh University Press.

Deleuze G and Guattari F (1980) *A Thousand Plateaus*. London and New York: Continuum.

Dépelteau F (2013) What Is the Direction of the "Relational Turn"? In: Powell C and Dépelteau F (eds) *Conceptualizing Relational Sociology*. New York: Palgrave Macmillan, pp. 163-185.

Doreian P and Conti N (2012) Social context, spatial structure and social network structure. *Social Networks* 34(1): 32-46.

Eveland WP, Appiah O and Beck PA (2018) Americans are more exposed to difference than we think: Capturing hidden exposure to political and racial difference. *Social Networks* 52(1): 192–200.

Everton SF (2012) *Disrupting dark networks*. Cambridge: Cambridge University Press.

Fleming L, King CK and Juda AI (2007) Small worlds and regional innovation. *Organization Science* 18(6): 938–954.

Flint C, Diehl P, Scheffran J, Vasquez J and Chi S-H (2009) Conceptualizing ConflictSpace: Toward a geography of relational power and embeddedness in the analysis of interstate conflict. *Annals of the Association of American Geographers* 99(5): 827–835.

Fouberg E, Murphy A and de Blij H (2015) *Human Geography: People, Place, and Culture*. New York: Wiley.

Frei A, Axhausen K and Ohnmacht T (2009) Mobilities and social network geography: Size and spatial dispersion-the Zurich case study. In: Maksim H, Ohnmacht T and Bergman M (eds) *Mobilities and Inequality*. Burlington, VT: Ashgate, pp. 99–120.

Grabher G (2006) Trading routes, bypasses, and risky intersections: mapping the travels of 'networks' between economic sociology and economic geography. *Progress in Human Geography* 30(2): 163–189.

Glückler J (2007) Economic geography and the evolution of networks. *Journal of Economic Geography* 7: 619–634.

Glückler J (2011) The importance of embeddedness in economic geography. *Geographische Zeitschrift* 89(4): 211–226.

Glückler J and Panitz R (2016) Unpacking social divisions of labor in markets: Generalized blockmodeling and the network boom in stock photography. *Social Networks* 47: 156-166.

Glückler J, Lazega E and Hammer I (2017). Exploring the interaction of space and networks in the creation of knowledge: An introduction. In: xxx (eds) *Knowledge and Networks*. Cham: Springer, pp. 1-21.

Glückler J and Doreian P (2016) Social network analysis and economic geography—positional, evolutionary and multi-level approaches. *Journal of Economic Geography* 16(6): 1123–1134.

Granovetter M (1983) The strength of weak ties: A network theory revisited. *Sociological Theory* 1: 201–233.

Granovetter M (1985) Economic action and social structure: The problem of embeddedness. *American Journal of Sociology* 91(3): 481–510.

Haggett P and Chorley RJ (1970) *Network Analysis in Geography*. New York: St Martin's Press.

Hadjimichalis C and Hudson R (2006) Networks regional development and democratic control. *International Journal of Urban and Regional Research* 30(4): 858–872.

Harris R (2009) Topology. In: Gregory D, Johnston R and Pratt G (eds) *The Dictionary of Human Geography*. Chichester: Wiley-Blackwell, p. 762.

Hess M (2004) Spatial relationships? Towards a reconceptualization of embeddedness. *Progress in Human Geography* 28(2): 165–186.

Hipp JR, Faris RW and Boessen A (2012) Measuring 'neighborhood': Constructing network neighborhoods. *Social Networks* 34(1): 128–140.

Howells J (2012) The geography of knowledge: Never so close but never so far apart. *Journal of Economic Geography* 12(5): 1003–1020.

Jones JP, Marston SA and Woodward K (2011) Scales and Networks Part II. In: Agnew JA and Duncan JS (eds) *The Wiley-Blackwell Companion to Human Geography*. Malden MA Wiley.

Juhász S and Lengyel B (2018) Creation and persistence of ties in cluster knowledge networks. *Journal of Economic Geography* 18: 1203-1226.

Kuhn TS (1963) *The Structure of Scientific Revolutions*. Chicago: University of Chicago Press.

Latour B (1993) *We Have Never Been Modern*. Cambridge MA: Harvard University Press.

Latour B (1996) On actor-network theory: A few clarifications. *Soziale Welt* 47: 369–381.

Latour B (2005) *Reassembling the Social: An Introduction to Actor-network-theory*. Oxford: Oxford University Press.

Leitner H (2004) The politics of scale and networks of spatial connectivity: transnational interurban networks and the rescaling of political governance in Europe. In: Sheppard E and McMaster RB (eds) *Scale and Geographic Inquiry: Nature, Society and Method*. Malden MA: Blackwell, pp. 236–255.

Maskell P (2014) Accessing remote knowledge—the roles of trade fairs pipelines crowdsourcing and listening posts. *Journal of Economic Geography* 14(5): 883-902.

Marshall DJ and Staeheli L (2015) Mapping civil society with social network analysis: Methodological possibilities and limitations. *Geoforum* 61: 56-66.

Mason MJ, Mennis J, Zaharakis N and Way T (2016) The dynamic role of urban neighborhood effects in a text-messaging adolescent smoking intervention. *Nicotine and Tobacco Research* 18(5): 1039–1045.

Massey D (1993) Power-geometry and a progressive sense of place. In: Bird J. Curtis B. Putnam T. Robertson G. Tickner L. (eds) *Mapping the Futures: Local Cultures Global Change*. London: Routledge, pp. 59–69.

Massey D (1994) *Space, Place and Gender*. Minneapolis MN: University of Minnesota Press.

McPherson M, Smith-Lovin L and Cook J (2001) Birds of a feather: homophily in social networks. *Annual Review of Sociology* 27: 415–44.

Menatti L (2013) A rhizome of landscapes: a geophilosophical perspective about contemporary global spaces. In: Newman C, Nussaume Y and Pedroli B (eds) *Landscape and Imagination: Towards a New Baseline for Education in a Changing World*. Florence: Bandecchi & Vivaldi, pp. 19–22.

Miller B (2000) *Geography and Social Movements: Comparing Antinuclear Activism in the Boston Area*. Minneapolis MN: University of Minnesota Press.

Mok D, Wellman B and Basu R (2007) Did distance matter before the Internet? Inter-personal contact and support in the 1970s. *Social Networks* 29 430–461.

Müller M and Schurr C (2016) Assemblage thinking and actor-network theory: conjunctions disjunctions cross-fertilisations. *Transactions of the Institute of British Geographers* 41(3): 217–229.

Neal ZP (2012) *The connected city: How networks are shaping the modern metropolis*. New York Routledge.

Nicholls WJ and Uitermark J (2017) *Cities and Social Movements: Immigrant Rights Activism in the US France and the Netherlands 1970-2015*. Malden MA: Wiley.

Onnela JP, Arbesman S, González MC, Barabási AL and Christakis NA (2011) Geographic constraints on social network groups. *PLoS ONE* 6(4): 1–7.

Paasi A (2011) Geography space and the re-emergence of topological thinking. *Dialogues in Human Geography* 1(3): 299–303.

Papachristos AV (2009) Murder by structure: Dominance relations and the social structure of gang homicide. *American Journal of Sociology* 115(1): 74–128.

Preciado P, Snijders TAB, Burk WJ, Stattin H and Kerr M (2012) Does proximity matter? Distance dependence of adolescent friendships. *Social Networks* 34(1): 18–31.

Radil SM, Flint C and Tita G (2010) Spatializing social networks: Using Social Network Analysis to investigate geographies of gang rivalry territoriality and violence in Los Angeles. *Annals of the Association of American Geographers* 100(2): 307–326.

Radil SM and Flint C (2013) Exiles and arms: the territorial practices of state making and war diffusion in post-Cold War Africa. *Territory Politics Governance* 1(2): 183–202.

Robbins P (2007) *Lawn People: How Grasses Weeds and Chemicals Make us Who we Are*. Philadelphia: Temple University Press.

Sailer K and McCulloh I (2012) Social networks and spatial configuration—How office layouts drive social interaction. *Social Networks* 34(1): 47–58.

Schaefer DR (2012) Youth co-offending networks: An investigation of social and spatial effects. *Social Networks* 34(1): 141-149.

Simonsen K (2004) Networks flows and fluids—reimagining spatial analysis? *Environment and Planning A* 36(8): 1333–1337.

Skillicorn D, Walther O, Zheng Q and Leuprecht C (2018) Spatial and temporal diffusion of political violence in North and West Africa. In: Walther O and Miles W (eds) *African Border Disorders*. London: Routledge, pp. 87–112.

Smith CJ (1980) Social networks as metaphors models and methods. *Progress in Geography* 4(4): 500–524.

Smith JM, Halgin DS, Kidwell-Lopez V, Labianca G, Brass DJ and Borgatti SP (2014) Power in politically charged networks. *Social Networks* 36 162–176.

Sohn C, Walther O and Christopoulos D (2010) The spatiality of social networks. Paper presented at the INSNA Sunbelt Conference July 1.

Staeheli LA (2003) Place. In: Agnew JA, Mitchell K and Toal G (eds) *A Companion to Political Geography*. Malden MA: Blackwell, pp. 158-170.

Takhteyev Y, Gruzd A and Wellman B (2012) Geography of Twitter networks. *Social Networks* 34(1): 73–81.

Tallman S and Phene A (2007) Leveraging knowledge across geographic boundaries. *Organization Science* 18(2): 252–260.

Taylor PJ (2004) *World City Network. A Global Urban Analysis*. London: Routledge.

Taylor PJ, Derudder B, Faulconbridge J, Hoyler M and Ni P (2014) Advanced producer service firms as strategic networks global cities as strategic places. *Economic Geography* 90(3): 267-291.

Ter Wal AL (2014) The dynamics of the inventor network in German biotechnology: geographic proximity versus triadic closure. *Journal of Economic Geography* 14(3): 589-

Walther O and Reitel B (2013) Cross-border policy networks in the Basel region: the effect of national borders and brokerage roles. *Space & Polity* 17(2): 217–236.

Wellman B (1979) The community question: The intimate networks of East Yorkers. *American Journal of Sociology* 84(5): 1201–1231.

White HC, Boorman SA and Breiger RL (1976) Social structure from multiple networks. I. Blockmodels of roles and positions. *American Journal of Sociology* 81(4): 730–780.